\documentstyle[preprint,tighten,aps,epsfig]{revtex}

\def\beq{\begin{equation}}
\def\eeq{\end{equation}}
\def\beqa{\begin{eqnarray}}
\def\eeqa{\end{eqnarray}}
\def\MeV{\nobreak\,\mbox{MeV}}
\def\GeV{\nobreak\,\mbox{GeV}}

\def\mb{\nobreak\,\mbox{mb}}

\def\mpi{m_\pi}
\def\md{m_D}
\def\mds{m_{D^*}}
\def\mpsi{m_\psi}
\begin{document}
\draft
\title{Charmonium - Pion Cross Section from QCD Sum Rules}
\author{F.S. Navarra$^1$, M. Nielsen$^1$,  R.S. Marques de Carvalho$^2$
and G. Krein$^2$}
\address{$^1$ Instituto de F\'{\i}sica, Universidade de S\~{a}o Paulo, 
C.P. 66318,  05315-970 S\~{a}o Paulo, SP, Brazil \\
$^2$ Instituto de F\'{\i}sica Te\'orica, Universidade Estadual Paulista,  
Rua Pamplona 145, 01405-900 S\~{a}o Paulo, SP, Brazil}

\maketitle
\begin{abstract}
The $J/\psi~\pi\rightarrow \bar{D}~D^*$, $D~\bar{D}^*$, ${\bar D}^*~D^*$ 
and ${\bar D}~D$ cross sections as a function of $\sqrt{s}$ are 
evaluated in a QCD sum rule calculation. We study the Borel sum rule 
for the four point function involving pseudoscalar and vector meson 
currents, up to dimension four in the operator product expansion. We find 
that our results are smaller than the  $J/\psi~\pi\rightarrow 
\mbox{charmed mesons}$ cross sections obtained with models based on 
meson exchange, but are close to those obtained with quark exchange 
models.
\end{abstract}
\narrowtext


Charmonium -  hadron cross sections are of crucial 
importance in the context of quark-gluon plasma  physics~\cite{bsw}. 
Small $J/\psi$ - hadron dissociation cross sections may favor an 
interpretation of the recent Pb + Pb data in terms of the production 
of a new phase of matter. Part of these interactions happens in the 
early stages of the nucleus - nucleus collisions and therefore at 
high energies ($\sqrt{s} \simeq 10 - 20$ GeV) and one may try to 
apply perturbative QCD. However, even in this regime, nonperturbative
effects may be important~\cite{dnn}. Interestingly, estimates using 
quite different methods give results clustering around the value of 
$3 - 5$~mb in this energy range. On the other hand, a significant 
part of the charmonium - hadron interactions occurs when other 
light particles have already been produced, forming a ``fireball''.
Interactions inside this fireball happen at much lower energies 
($\sqrt{s} \le 5$ GeV) and one has to apply nonperturbative methods.

One possible nonperturbative reaction mechanism is meson exchange, 
which can be studied by means of effective Lagrangians, constrained 
by flavor and chiral symmetries as well as by gauge invariance. 
This approach was first introduced in ref.~\cite{mamu98} and further
developed by other groups~\cite{osl,haglin,linko,nnr}. Another 
reaction mechanism is quark interchange driven by Born-order matrix
elements of the standard nonrelativistic quark model~\cite{hksv,wongs,mbq}. 
In this approach, once the masses and sizes of the mesons are fixed, 
there are no free parameters left.

The results of the calculations for the charmonium-pion cross sections
based on these two approaches can differ by two orders of magnitude in 
the relevant energy range. The situation clearly calls for different 
types of calculations that are constrained by other, independent pieces 
of phenomenology. In this work we use the QCD sum rules (QCDSR)
technique~\cite{svz,rry} to study the $J/\psi-\pi$ dissociation. The 
QCDSR technique allows one to compute hadronic quantities like masses, 
coupling constants and form factors in terms of quark and gluon 
properties and universal matrix elements which represent the properties 
of the QCD vacuum.  In view of our relatively poor understanding of
$J/\psi$ reactions in nuclear matter and considering the large
discrepancies between different model estimates, we believe that our 
work adds to a better understanding of this important topic.

We consider all four channels $J/\psi~\pi \rightarrow$:
$\bar{D}~D^*$, $D~\bar{D}^*$, $\bar{D}~D$ and $\bar{D}^*~D^*$.
Let us start with the the four-point function for the process 
$J/\psi~\pi\rightarrow \bar{D}~D^*$:
\beqa
\Pi_{\mu\nu} &=& i\int d^4x~d^4y~d^4z~e^{-ip_1.x}~e^{ip_3.y}~
e^{ip_4.z} \nonumber\\
&\times&
\langle 0|T\{j _{\pi}(x)j_\nu^{D^*}(y)j_\mu^\psi(0)j_D(z)\}|0\rangle \;, 
\label{cor}
\eeqa
with the currents given by $j_\pi=\overline{d} i\gamma_5u$, 
$j_\nu^{D^*}=\overline{u} \gamma_\nu c $,
$j_\mu^\psi=\overline{c} \gamma_\mu c$ and
$j_D=\overline{c} i\gamma_5d$ \cite{rry},
where $c$, $u$ and $d$  are the charm, up and down quark fields
respectively, and $p_1$, $p_2$, $p_3$ and $p_4$ are the four-momenta of the 
mesons $\pi$, $J/\psi$, $D^*$ and $D$ respectively, with $p_1+p_2=p_3+p_4$.

The phenomenological side of the correlation function, $\Pi_{\mu\nu}$,
is obtained by the consideration of $J/\psi$, $\pi$, $D$ and $D^*$ state 
contribution to the matrix element in Eq.~(\ref{cor}):
\beqa
&&\Pi_{\mu\nu}^{phen} = -{\mpi^2 F_\pi\over m_u+m_d}{\md^2 f_D\over m_c}~
{\mds f_{D^*}~\mpsi f_\psi~{\cal{M}}^{\alpha\beta}\over
(p_1^2-\mpi^2)(p_4^2-\md^2)} \nonumber\\
&&\,\times\,{g_{\mu\alpha}-p_{2\mu}p_{2\alpha}/\mpsi^2\over
p_2^2-\mpsi^2}~{g_{\nu\beta}-p_{3\nu}p_{3\beta}/\mds^2\over
p_3^2-\mds^2} + \mbox{h. r.}\; ,
\label{phen}
\eeqa
where h. r. means higher resonances and the hadronic amplitude for the 
process $J/\psi~\pi\rightarrow \bar{D}~D^*$ is given by 
\beq
{\cal M} = {\cal{M}}_{\mu\nu}(p_1,p_2,p_3,p_4)~\epsilon_2^{\mu}
~\epsilon_3^{*\nu}\,.
\label{calM}
\eeq

We note that one has $1/p_1^2$ pole in Eq.~(\ref{phen}) in the limit of 
a vanishing pion mass. Following~\cite{ccl,bnn,nos,rry}, we can write 
a sum rule at $p_1^2=0$ and single out the leading terms in the operator
product expansion (OPE) of Eq.~(\ref{cor}) that match the $1/p_1^2$ term. 
The perturbative diagram does not contribute with $1/p_1^2$ and, up to 
dimension four, only the diagrams proportional to the quark condensate, 
shown in Fig.~1, contribute. After collecting the $1/p_1^2$ terms on 
the theoretical side and taking the limit $p_{1\mu}\rightarrow0$ in the 
residue of the pion pole, one obtains for the contribution of these two 
graphs
\beq
\Pi_{\mu\nu}^{<\bar{q}q>}=-{2m_c\langle\bar{q}q\rangle\over p_1^2}
{ p_{1\nu} ( p_{1\mu} + p_{2\mu} - 2 p_{3\mu} ) - p_{1\mu} p_{2\nu} 
\over(p_3^2-m_c^2)(p_4^2-m_c^2)}\; .
\label{qbarq}
\eeq

Contracting the hadronic amplitude with the numerators of $J/\psi$ and 
$D^*$ propagators in Eq.~(\ref{phen}) and comparing with Eq.~(\ref{qbarq}), 
the structure defining ${\cal{M}_{\mu\nu}}$ in Eq.~(\ref{calM}) is
easily identified. Therefore, defining
\beq
{\cal{M}}_{\mu\nu}=\Lambda_{DD^*}~\left(p_{1\mu}p_{1\nu}-p_{1\mu}p_{2\nu}
-2p_{1\nu}p_{3\mu}\right)\;,
\label{mmunu}
\eeq
we can write a sum rule for $\Lambda_{DD^*}$ in any of the three
structures appearing in Eq.~(\ref{mmunu}). To improve the matching between
the phenomenological and theoretical sides we follow the usual procedure
and make a single Borel transform, with all the external momenta (except 
$p_1^2$) taken to be equal: $-p_2^2=-p_3^2=-p_4^2=P^2\rightarrow M^2$. 
The problem of doing a single Borel transformation is the fact that terms 
associated with the pole-continuum  transitions are not
suppressed~\cite{io2}. In ref.~\cite{io2} it was explicitly shown that 
the pole-continuum transition has a different behavior as a function of 
the Borel mass as compared with the double pole contribution (triple pole
contribution in our case) and continuum contribution: it grows with $M^2$
as compared with the contribution of the fundamental states. Therefore, 
the pole-continuum contribution can be taken into account through the
introduction of a parameter $A_{DD^*}$ in the phenomenological side of 
the sum rule~\cite{bnn,nos,io2}. Thus, neglecting $m_\pi^2$ in the 
denominator of Eq.~(\ref{phen}) and doing a single Borel transform in
$-p_2^2=-p_3^2=-p_4^2=P^2$, we get 
\beqa
&& {\Lambda_{DD^*}+A_{DD^*}M^2\over\mds^2-\mpsi^2}\Biggl[{e^{-\md^2/M^2}-
e^{-\mpsi^2/M^2} \over\mpsi^2-\md^2}~- (\psi\rightarrow D^*)\Biggr]
\nonumber\\
&&= - 2m_c\langle\bar{q}q\rangle~
{e^{-m_c^2/M^2}\over M^2}  {m_c(m_u+m_d)\over
\mpi^2\md^2\mds\mpsi F_\pi f_Df_{D^*}f_\psi}\,,
\label{sr}
\eeqa
where we have transferred to the theoretical side the couplings of the
currents with the mesons, and have introduced, in the phenomenological
side, the parameter $A_{DD^*}$ to account for possible nondiagonal transitions.

At this point we should mention that the approximations we are using
of exploiting the soft-pion limit and making the single Borel transform
presents uncertainties. The main uncertainties are related to the 
continuum subtraction, the non-diagonal contributions, and the
subtraction terms in the multiple dispersion relation. The approximation
of the soft-pion limit can be ameliorated by going off the pion pole. 
In addition, further improvements can be made by use of light-cone 
sum-rules~\cite{LCSR}. These allow to use a light-cone pion distribution 
amplitude in substitution of condensates incorporating in this way 
additional QCD effects. Nevertheless, we believe that in this initial 
attempt our results should be useful as a comparison with what is obtained 
using model calculations.

For consistency we use in our analysis the QCDSR expressions for 
the decay constants of the $J/\psi,~D^*$ and $D$ mesons
up to dimension four in lowest order 
of $\alpha_s$:
\beqa
&&f_D^2 = {3m_c^2\over 8\pi^2m_D^4}\int_{m_c^2}^{u_D}du~ 
{(u-m_c^2)^2\over u}e^{(m_D^2-u)/M_M^2}\,
\nonumber\\
&&\,- {m_c^3\over m_D^4}
\langle\bar{q}q\rangle e^{(m_D^2-m_c^2)/M_M^2}\; ,\label{fhr} \\
&&f_{D^*}^2 = {1\over 8\pi^2\mds^2}\int_{m_c^2}^{u_{D^*}}ds~ 
\Biggl[{(s-m_c^2)^2
\over s}
\left(2+{m_c^2\over s}\right) \nonumber\\
&&\,\times\, e^{(\mds^2-s)/M_M^2}\Biggr]
\,-\,{ m_c\over \mds^2}\langle\bar{q}q\rangle e^{(\mds^2-m_c^2)/M_M^2},
\label{fhs} \\
&&f_\psi^2 = {1\over4\pi^2}\int_{4m_c^2}^{u_\psi}dr~{(r+2m_c^2)
\sqrt{r-4m_c^2}\over r^{3/2}}e^{(\mpsi^2-r)/M_M^2},\label{fpsi}
\eeqa
where $M_M^2$ represents the Borel mass in the two-point function. We 
have also omitted the  numerically insignificant contribution of the
gluon condensate.

The parameter values used in all calculations are $m_u+m_d=14\,\MeV$, 
$m_c=1.5\,\GeV$, $m_\pi=140\,\MeV$, $m_D=1.87\,\GeV$, $m_{D^*}=2.01\,
\GeV$, $\mpsi=3.097\,\GeV$, $F_\pi=\sqrt{2}f_\pi=131.5\,\MeV$,
$\langle\overline{q}q\rangle\,=\,-(0.23)^3\,\GeV^3$.
We parametrize the continuum thresholds as $u_M=(m_{M}+\Delta_u)^2$. 
The values of $u_M$ are, in general, extracted from the two-point
function sum rules for $f_D$ and $f_{D^*}$ and $f_\psi$ in Eqs.~(\ref{fhr}), 
(\ref{fhs}) and (\ref{fpsi}). Using the Borel region 
$3 \leq M_M^2\leq 6 \GeV^2$ for the $D^*$ and  $D$ mesons and 
$6 \leq M_M^2 \leq 10 \GeV^2$ for the $J/\psi$, we found good stability 
for $f_D$,  $f_{D*}$ and $f_\psi$ with $\Delta_u\sim0.6\GeV$. We obtained 
$f_D=155\pm5\MeV$, $f_{D^*}=195\pm5\MeV$ and $f_\psi=225\pm10\MeV$, which
are acceptable values for these decay constants \cite{fs}. 
However, instead of using numerical values for these decay constants we 
use the sum rules in Eqs.~(\ref{fhr}), (\ref{fhs})
and (\ref{fpsi}) directly when evaluating ${\cal{M}}$.

In Ref.~\cite{bbg} it was found that relating the Borel parameters in 
the two-and three-point functions through $M^2=2~M_M^2$, is a crucial
ingredient for the incorporation of heavy quark symmetries, and leads 
to a considerable reduction of the sensitivity to input parameters,
such as the continuum thresholds, and to radiative corrections. Therefore,
we will use  $M^2=2~M_M^2$ to relate the Borel parameters and will work
in the Borel range $8 \leq M^2 \leq 16 \GeV^2$. We recall that this region
corresponds to $4 \leq M_M^2 \leq 8 \GeV^2$, in which we have obtained 
good stability for the two-point sum rules of Eqs.~(\ref{fhr}), 
(\ref{fhs}) and (\ref{fpsi}). This region also covers the range of the
average values of the masses of the $D$, $D^*$ and $J/\psi$ mesons.   

In Fig.~2 we show, for $\Delta_u=0.6\,\GeV$, the QCD sum rule results for 
$\Lambda_{DD^*}+A_{DD^*}M^2$ as a function of $M^2$ (dots). We see that
they follow a straight line in the Borel region $8\leq M^2\leq16\,\GeV^2$.
The value of the amplitude $\Lambda$ is obtained by the extrapolation of
the line to $M^2=0$~\cite{bnn,nos,io2}. Fitting the QCD sum rule results
to a straight line we get
\beq
\Lambda_{DD^*}\simeq17.71\GeV^{-2}\;.
\label{ampli}
\eeq
As expected, in our approach $\Lambda$ is just a number and all dependence 
of ${\cal M}_{\mu\nu}$ (Eq.~(\ref{mmunu})) on particle momenta is contained 
in the Dirac structure. This is a consequence of our low energy
approximation.

Next, we consider the process $J/\psi~\pi\rightarrow \bar{D}~D$ ($J/\psi~\pi
\rightarrow \bar{D}^*~D^*$). In this case we have to change the current 
$j_\nu^{D^*}$ ($j_D$) in Eq.~(\ref{cor}) to $\bar{u}i\gamma_5c$ 
($\bar{c}\gamma_\alpha d$). The phenomenological side is obtained as
\beqa
&&\Pi_{\mu}^{phen} = -{\mpi^2 F_\pi\over m_u+m_d}{\md^4 f_D^2\over m_c^2}
\,{-g_{\alpha\mu}+p_{2\alpha}p_{2\mu}\over(p_2^2-\mpsi^2)}\nonumber\\
&&\,\times\,{\mpsi f_\psi~{\cal M}^\alpha\over(p_1^2-\mpi^2)(p_3^2-\md^2)
(p_4^2-\md^2)} + \mbox{h. r.}\,,
\label{phendd}
\eeqa
for $J/\psi~\pi\rightarrow \bar{D}~D$, where the hadronic amplitude 
is defined by ${\cal M}={\cal{M}}_{\mu}(p_1,p_2,p_3,p_4)~\epsilon_2^{\mu}$.
In the same way, for $J/\psi~\pi\rightarrow \bar{D}^*~D^*$ we get
\beqa
&& \Pi_{\mu\nu\alpha}^{phen}=-{\mpi^2 F_\pi\over m_u+m_d}~
{\mds^2 f_{D^*}^2~\mpsi f_\psi~{\cal{M}}^{\beta\delta\sigma}\over
(p_1^2-\mpi^2)}\nonumber\\
&&\times\, {-g_{\mu\beta}+p_{2\mu}p_{2\beta}/\mpsi^2\over
p_2^2-\mpsi^2}\,{-g_{\nu\delta}+p_{3\nu}p_{3\delta}/\mds^2\over
p_3^2-\mds^2}\nonumber\\
&&\times\,{-g_{\alpha\sigma}+p_{4\alpha}p_{4\sigma}/\mds^2\over
p_4^2-\mds^2} +\mbox{h. r.}\,,
\label{phendsds}
\eeqa
with the corresponding hadronic amplitude defined by ${\cal M}=
{\cal{M}}_{\mu\nu\alpha}(p_1,p_2,p_3,p_4)~\epsilon_2^{\mu}~\epsilon_3^{*\nu}
~\epsilon_4^{*\alpha}$.

Similarly to the case $J/\psi~\pi\rightarrow \bar{D}~D^*$,
in the OPE side the only diagrams, up to dimension four, contributing
with $1/p_1^2$ are the diagrams shown in Fig.~1. Therefore, taking the limit $p_{1\mu}\rightarrow0$ in the residue of the pion pole we get:
\beq
\Pi_{\mu}^{<\bar{q}q>}=-{2\langle\bar{q}q\rangle\over p_1^2}{
\epsilon_{\mu\alpha\beta\sigma}p_{1}^\alpha p_3^\beta p_4^\sigma 
\over(p_3^2-m_c^2)(p_4^2-m_c^2)}\; ,
\label{qqdd}
\eeq
and
\beqa
\Pi_{\mu\nu\alpha}^{<\bar{q}q>}&=&-{2\langle\bar{q}q\rangle\over p_1^2
(p_3^2-m_c^2)(p_4^2-m_c^2)} \nonumber\\
&\times& \left[(m_c^2 + p_3.p_4)\, \epsilon_{\alpha\mu\nu\beta} \,p_1^\beta 
+  E_{\mu\nu\alpha}\right]\,,
\label{qqdsds}
\eeqa
where
\beqa
&& E_{\mu\nu\alpha} =
p_1^\beta p_3^\lambda p_4^\gamma(-\epsilon_{\nu\beta\lambda\gamma}
g_{\alpha\mu}+\epsilon_{\mu\beta\lambda\gamma}g_{\alpha\nu}
-\epsilon_{\alpha\beta\lambda\gamma}g_{\mu\nu}) \nonumber\\
&& + \, \epsilon_{\mu\nu\beta\lambda}(p_1^\beta p_4^\lambda p_{3\alpha}-
p_3^\beta p_4^\lambda p_{1\alpha})+\epsilon_{\alpha\nu\beta\lambda}
(p_1^\beta p_3^\lambda p_{1\mu} \nonumber\\
&& + \, p_3^\beta p_4^\lambda p_{1\mu}
-p_1^\beta p_4^\lambda p_{3\mu}-p_1^\beta p_3^\lambda p_{4\mu})
+\epsilon_{\alpha\mu\beta\lambda}(-p_1^\beta p_3^\lambda p_{1\nu}
\nonumber\\
&& - \, p_3^\beta 
p_4^\lambda p_{1\nu}+p_1^\beta p_4^\lambda p_{1\nu}+p_1^\beta p_3^\lambda 
p_{4\mu}) \,.
\label{est}
\eeqa

Comparing the phenomenological and OPE sides of the correlators we can 
identify the structure defining the hadronic amplitudes:
\beq
{\cal{M}}_{\mu}=\Lambda_{DD}~\epsilon_{\mu\alpha\beta\sigma}p_{1}^\alpha 
 p_3^\beta p_4^\sigma\,,\;\;\;{\cal{M}}_{\mu\nu\alpha}=
\Lambda_{D^*D^*}~E_{\mu\nu\alpha}\,.
\label{ampmm}
\eeq

It is important to notice that in writing Eq.~(\ref{ampmm}) we  have
neglected the structure $\epsilon_{\alpha\mu\nu\beta}p_1^\beta$ in 
${\cal{M}}_{\mu\nu\alpha}$. This is because,
as can be seen from Eq.~(\ref{qqdsds}), this structure contains a term
$p_3{\cdot}p_4$ that can be rewritten in terms of $p_3^2-m_c^2$ and 
$p_4^2-m_c^2$ and, therefore, will contribute with a single pole which 
contains information about pole-excited states contributions. Since these 
contributions are considered in the phenomenological side as a parameter, 
we do not need to include them explicitly in the OPE side.

We can write a sum rule for $\Lambda_{DD}$ in the structure 
$\epsilon_{\mu\alpha\beta\sigma}p_{1}^\alpha  p_3^\beta p_4^\sigma$, 
and a sum rule for $\Lambda_{D^*D^*}$ in any of the structures appearing
in Eq.~(\ref{est}). Thus, neglecting $m_\pi^2$ in the denominator of
Eqs.~(\ref{phendd}) and (\ref{phendsds}), and doing a single Borel
transform in $-p_2^2=-p_3^2=-p_4^2=P^2$, we get 
\beqa
&&{\Lambda_{MM}+A_{MM}M^2\over m_M^2-\mpsi^2} \, f_M(M^2) =
C_M \, {m_u+m_d\over\mpi^2 m_M^2\mpsi F_\pi f_{M}^2f_\psi}\nonumber\\
&&\times\, 2\,\langle\bar{q}q\rangle {e^{-m_c^2/M^2}\over M^2}\,,
\label{srmm}
\eeqa
where the subscript $M$ stands for the $D$ or $D^*$ mesons, with $C_D={m_c^2
\over\md^2}$, $C_{D^*}=1$ and
\beq
f_M(M^2)={e^{-m_M^2/M^2}
\over M^2}-{e^{-m_M^2/M^2}-e^{-\mpsi^2/M^2}
\over\mpsi^2-m_M^2}\;.
\eeq

In Fig.~2 we also show, for $\Delta_u=0.6\,\GeV$, the QCD
sum rule results for  $\Lambda_{DD}+A_{DD}M^2$ (diamonds) and 
$\Lambda_{D^*D^*}+A_{D^*D^*}M^2$ (triangles)
as a function of $M^2$  from where we see that, in the Borel region
$8\leq M^2\leq16\,\GeV^2$, they all follow a straight line. As explained
before, the value of the 
amplitudes $\Lambda_{DD}$ and $\Lambda_{D^*D^*}$ are obtained by the 
extrapolation of the line to $M^2=0$. We get:
\beq
\Lambda_{DD}\simeq12.25\GeV^{-1}\;,\;\;\;\Lambda_{D^*D^*}\simeq11.39\GeV^{-3}
\;.\label{amplim}
\eeq


Having the QCD sum rule results for the amplitude of the three processes
$J/\psi~\pi\rightarrow \bar{D}~D^*,~\bar{D}~D,~\bar{D}^*~D^*$, given in 
Eqs.~(\ref{mmunu}) and (\ref{ampmm})
with $\Lambda$ given in Eqs.~(\ref{ampli}) and (\ref{amplim})
we can evaluate  the differential cross section.
%

Using our QCD sum rule result in Eqs.~(\ref{mmunu}), (\ref{ampmm}), 
(\ref{ampli}) and  (\ref{amplim}) we show, in Fig.~3,  the cross section 
for the $J/\psi~\pi$ dissociation. It is important to keep in mind that,
since our sum rule was derived in the limit $p_{1}\rightarrow0$,
we can not extend our results to large values of $\sqrt{s}$. Also,
since the perturbative contribution is absent in our calculation, 
we were not able to properly disentangle the continuum contribution and 
our cross section may include contributions from higher states. Whereas
they are certainly not important in the case of the pion, they may give
some contribution to the heavy currents. Therefore our $J/\psi~\pi$ cross
section may implicitly include (at least partially) the process
$\psi'~\pi$. For this reason, our numbers might be regarded as upper
bounds. 

Our first conclusion is that our results show that, for  values
of $\sqrt{s}$ far from the $J/\psi~\pi\rightarrow \bar{D}^{*}~{D}^*$
threshold,  $\sigma_{J/\psi\pi\rightarrow \bar{D}^{*}{D}^*} \, \geq \,
\sigma_{J/\psi\pi\rightarrow \bar{D}{D}^*+D\bar{D}^*} \, \geq \,
\sigma_{J/\psi\pi\rightarrow \bar{D}{D}}$, in agreement with the model 
calculations presented in \cite{osl} but in disagreement with 
the results obtained with the nonrelativistic quark model of \cite{wongs}, 
which show that the state $\bar{D}^*D$ has a larger production cross
section than $\bar{D}^{*}{D}^*$. Furthermore, our curves indicate that 
the cross section grows monotonically with the c.m.s. energy but not as fast, 
near the thresholds, as it does in the calculations in
Refs.~\cite{osl,haglin,linko,nnr}. Again, this behavior is in
opposition to~\cite{wongs}, where a peak just after the threshold 
followed by continuous decrease in the cross section was found.   

At higher energies, due to our low energy approximation, our approach 
gradually loses validity. In the fiducial region, close to threshold, 
$4.1 \, \leq \, \sqrt{s} \, \leq \, 4.3 \, \GeV$, 
we find $2.5 \, \leq \, \sigma \, \leq \, 4.0 \, \mb$ and these values 
are much smaller than those obtained with the effective Lagrangians 
without form factors in the hadronic vertices, but agree in order of
magnitude with the quark model calculations  of~\cite{wongs}. 

Finally, we should mention that we have been studying the dissociation 
processes of the $J/\psi$ in vacuum and the quantities relevant for QGP 
physics are in medium cross sections. In our approach the main effect 
introduced by the medium is the modification of the condensates, which is 
thought to be very mild. Our results depend only on the quark condensate 
and since it decreases with the nuclear density, we expect a further 
reduction in our cross section in a dense nuclear environment. 
 
In conclusion, we have used the QCD sum rule approach to evaluate the 
hadronic amplitude of the $J/\psi~\pi$ dissociation. From the hadronic 
amplitude we have evaluated the $J/\psi~\pi\rightarrow$ charmed mesons
dissociation cross section, and have obtained 
$2.5 \, \leq \, \sigma \, \leq \, 4.0 \, \mb$
at $4.1 \, \leq \, \sqrt{s} \, \leq \, 4.3 \, \GeV$. In view of the 
uncertainties discussed above these numbers should be taken as upper 
limits.

It is interesting to remember that Bhanot and Peskin \cite{bhp} have also 
used the 
OPE in the short distance limit to study the charmonium hadron
cross section. This work was latter enlarged and updated by Kharzeev et al.
\cite{kha} and also by Oh et al. \cite{okl}. In these papers the crucial 
assumption was
made that the charmonium is very small and resolves the partonic structure
of the light hadron. In our approach we do not use this assumption, and we
obtain larger values for the cross section. This seems to indicate that
size effects are important, and that the $J/\psi$ cannot be considered as 
a nearly point like object.

\vspace{0.5cm}
This work was supported by CNPq and FAPESP (contract numbers 
98/06590-2, 99/12987-5 and 00/04422-7).

   


\begin{figure}
\begin{center}
\epsfig{file=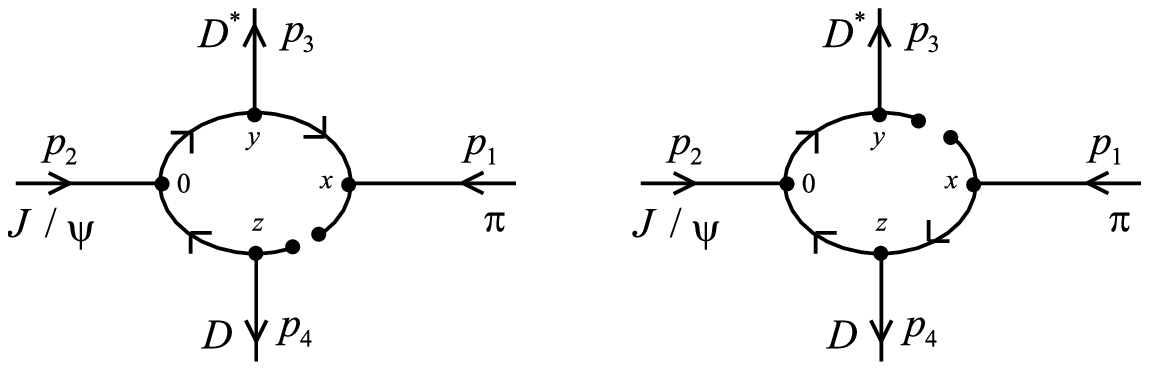,height=2.9cm}
\caption{Diagrams that contribute with $1/p_1^2$, up to dimension four,
in the OPE side of the amplitude $\pi+J/\psi\rightarrow D+D^*$.}
\label{fig1}
\end{center}
\end{figure}

\begin{figure}
\begin{center}
\epsfig{file=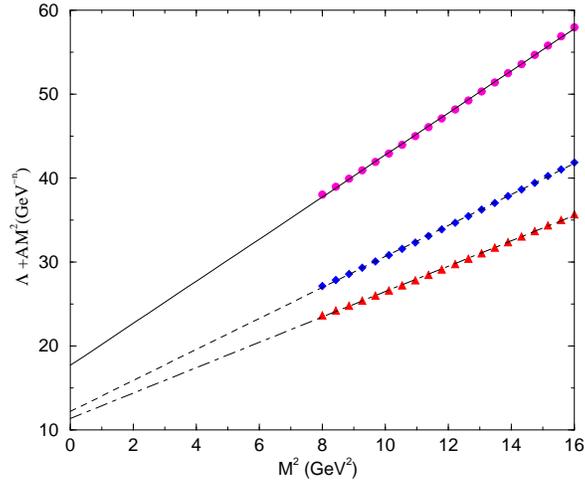,height=6.5cm}
\caption{Amplitudes of the processes $\pi~J/\psi\rightarrow$ 
$\bar{D}~D^* + D\bar{D}^*$ (dots), $\bar{D}~D$ (diamonds) and 
$\bar{D}^* D^*$ (triangles) as a function of the squared Borel mass
$M^2$. The solid, dotted and dot-dashed lines give the extrapolations to 
$M^2=0$ (respectively). }
\label{fig2}
\end{center}
\end{figure}

\begin{figure} 
\begin{center}
\epsfig{file=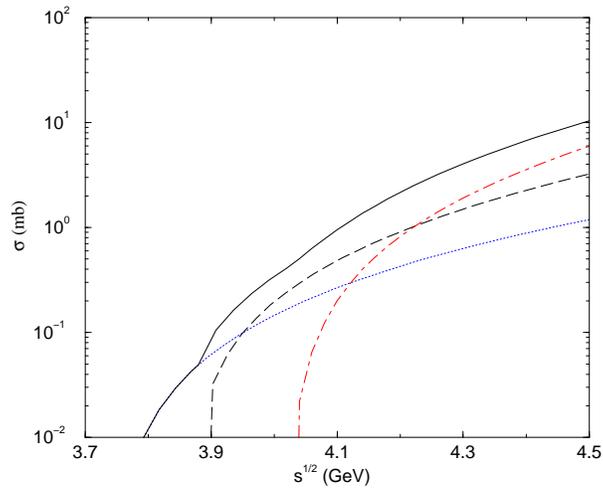,height=6.5cm}
\caption{Total cross sections of the processes $J/\psi~\pi\rightarrow$
$\bar{D}~D^* + D~\bar{D}^*$ (dashed line), $\bar{D}~D$ 
(dotted line) and $\bar{D}^*~D^*$ (dot-dashed line). 
The solid line gives the total $J/\psi~\pi$ dissociation cross section.}
\label{fig3}
\end{center}
\end{figure}

\end{document}